\documentclass[notitlepage, 12pt]{article}
\usepackage[margin=1in]{geometry}
\usepackage{verbatim,amsmath,graphicx,amssymb,amsfonts,amsthm,enumerate,setspace,natbib,hyperref,booktabs,siunitx,caption,threeparttable,lmodern,microtype,epstopdf,subcaption}
\let\estinput=\input

\newcommand{\estauto}[3]{
		\vspace{.75ex}{
			\begin{tabular}{l*{#2}{#3}}
			\toprule
			\estinput{#1}
			\bottomrule
			\addlinespace[.75ex]
			\end{tabular}
			}
		}

\newcommand{\figtext}[1]{
	\captionsetup{justification=justified,font=footnotesize}
	\caption*{\hspace{6pt}\hangindent=1.5em #1}
	}

\newcommand{\starnote}{\figtext{* $p < 0.1$, ** $p < 0.05$, *** $p < 0.01$. Standard errors in parentheses.}}

\title{The effect of stock market indexing on corporate tax avoidance}
\author{Alex Young \\ North Dakota State University}

\begin{document}
\maketitle
\begin{abstract}
\noindent Membership in the Russell 1000 and 2000 Indices is based on a ranking of market capitalization in May. Each index is separately value weighted such that firms just inside the Russell 2000 are comparable in size to firms just outside (i.e.\ at the bottom of the Russell 1000) but have much higher index weights. These features allow for the the annual reconstitution of these indices to be used as part of a regression discontinuity design to identify the effect of stock market indexing. Using this design, I investigate whether stock market indexing affects corporate tax avoidance. I find no evidence that firms just inside the Russell 2000 have significantly different effective tax rates than firms just outside.
\end{abstract}
\footnotetext[1]{E-mail: \texttt{alex.young.1@ndsu.edu}.} 
\newpage
\section{Introduction}
The ``file drawer'' problem refers to the practice of researchers abandoning results that they believe journals are unlikely to publish \citep{mervis:2015}. The belief is supported by evidence of publication bias in, for example, the social sciences: studies with results that reject the specified null hypothesis are forty percentage points more likely to be published than studies that fail to reject the null hypothesis (hereafter ``null results'') and are sixty percentage points more likely to be written up \citep{fms:2014}. There is some justification for this publication bias. One does not accept a null hypothesis; one fails to reject a null hypothesis \citep{hayashi:2000}. Furthermore, as stated by political scientist Gary King, ``there are many, many ways to produce null results by messing up'' \citep{mervis:2015}.
\newline

\noindent Anecdotally, my experience is consistent with the results of \citet{fms:2014}: I began this project in May 2014, found a null result, and didn't bother to write anything up. But really, the ``file drawer'' problem is stupid, so in this paper, I investigate and fail to reject the hypothesis that stock market indexing affects corporate tax avoidance. \textit{Ex ante}, a credible null hypothesis is that stock market indexing has no effect on corporate tax avoidance because passive indexers adjust their portfolios mechnically as the underlying index constituents change; thus, they hold firms regardless of the level of tax avoidance. However, to the extent that tax avoidance improves firm value and passive indexers are not passive owners, stock market indexing may increase tax avoidance.

\section{Empirical Framework}\label{researchdesign}
\subsection{Data}
The data on Russell 1000 and 2000 Index constituents come from FTSE Russell.\footnote{I thank Harrison Hong for providing the data.} The sample period covers the annual reconstitutions from 1996 to 2012. To compute the measures of corporate tax avoidance, I use data from Compustat.

\subsubsection{Identification Strategy}
As described in \citet{chl:2015}, the process of Russell 1000 and 2000 Index reconstitution generates exogenous variation in stock market indexing. On the last trading day in May, firms are ranked by their market capitalizations. Prior to 2006, the 1000 largest firms were assigned to the Russell 1000, and the next 2000 firms were assigned to the Russell 2000.\footnote{After 2006, Russell implemented a ``banding'' policy to reduce turnover between the indices. The practical consequence was that there became two cutoffs: one for firms entering the Russell 2000 from the Russell 1000, and another for firms entering the Russell 1000 from the Russell 2000.} FTSE Russell then publishes index weights in the end of June based on a measure of market capitalization that adjusts for publically nontradeable shares.
\newline

\noindent Each index is separately value weighted such that firms at the top of the Russell 2000 have significantly higher index weights than firms at the bottom of the Russell 1000, despite being similar in size. Thus, small variations in market capitalization at the end of May have large, discontinuous effects on index weights at the end of June. The process of index reconstitution thus lends itself to a fuzzy regression discontinuity (RD) design. The fuzziness arises because researcher-constructed rankings of May market capitalization do not perfectly predict index assignment.\footnote{Note that using rankings imputed from June index weights is inappropriate under any circumstance, as it results in a discontinuity in May market capitalization at the cutoff for Russell 2000 Index membership \citep{chl:2015}.}

\subsection{Research Design}
Following \citet{chl:2015}, I implement the fuzzy regression discontinuity design by estimating the following with two-stage least squares (2SLS):
\begin{align}\label{eq:fuzzy}
\text{R2000}_{i,t} &= \alpha_{0} + \alpha_{1}\tau_{i,t} + \alpha_{2}\text{Rank}_{i,t} + \alpha_{3}\tau_{i,t}\times\text{Rank}_{i,t} + \nu_{i,t} \notag \\
\text{ETR}_{i,t} &= \beta_{0} + \beta_{1}\text{R2000}_{i,t} + \beta_{2}\text{Rank}_{i,t} + \beta_{3}\text{R2000}_{i,t}\times\text{Rank}_{i,t} + \varepsilon_{i,t}
\end{align}

\noindent where $\tau_{i,t}$ is an indicator variable equal to 1 if firm $i$ crosses the threshold for Russell 2000 Index membership in year $t$ and is the instrumental variable for actual index assignment, $\text{R2000}_{i,t}$. Standard errors are adjusted for heteroskedasticity, and the bandwidth is 200.
\newline

\noindent $\text{ETR}_{i,t}$ is either the cash effective tax rate (CETR) or the GAAP effective tax rate (GETR) following \citep{dhm:2010}:
\begin{itemize}
\item $\text{CETR}_{i,t}$ is defined as cash taxes paid (Compustat TXPD) divided by pretax income less special items (Compustat PI - SPI).

\item $\text{GETR}_{i,t}$ is defined as total tax expense (Compustat TXT) divided by pretax income less special items (PI - SPI).
\end{itemize}

\noindent If the denominator is zero, then the ETR variables are set to missing. If the ETR variables are greater than 1 (less than 0), they are set to 1 (0).

\section{Results}\label{results}
Table \ref{table:iortax} presents the results of estimating Eq. \eqref{eq:fuzzy}. Column 1 shows that firms that cross the addition threshold are 87\% more likely to be added to the Russell 2000 than firms that stay in the Russell 1000. Column 2 shows that firms that cross the deletion threshold are 86.1\% more likely to be deleted from the Russell 2000 compared to firms that stay in the Russell 2000. The coefficients on the instrument, $\tau_{i,t}$, are significantly different from zero. In untabulated results, I find that the $F$-statistics testing the null of instrument insignificance easily pass the \citet{ss:1997} critical value. Thus, there is no evidence that the instrument is weak.
\newline

\noindent Columns 3 through 6 show that there is no evidence that firms just inside the Russell 2000 have significantly different cash or GAAP ETRs than firms just outside the Russell 2000 (i.e.\ at the bottom of the Russell 1000). Thus, I find no evidence that stock market indexing affects corporate tax avoidance.

\section{Conclusion}\label{conclusion}
In this paper, I investigate whether stock market indexing affects corporate tax avoidance. Using the annual reconstitution of the Russell 1000 and 2000 Indices in a fuzzy regression discontinuity design, I find no evidence that stock market indexing affects cash or GAAP effective tax rates.

\newpage
\bibliography{master}
\bibliographystyle{abbrvnat}
\newpage
\begin{table}
\caption{The effect of stock market indexing on corporate tax avoidance}\centering
\estauto{iortax2}{6}{c}\\
\starnote
\figtext{This table reports the results of a fuzzy RD design from the following equation. The dependent variable in columns 1 and 2 is an indicator for membership in the Russell 2000 Index. In columns 3 and 4 (5 and 6), the dependent variable is the cash (GAAP) effective tax rate Standard errors are adjusted for heteroskedasticity. The bandwidth is 200. The sample period is 1996--2012.}
\footnotesize\begin{align*}
\text{R2000}_{i,t} &= \alpha_{0} + \alpha_{1}\tau_{i,t} + \alpha_{2}\text{Rank}_{i,t} + \alpha_{3}\tau_{i,t}\times\text{Rank}_{i,t} + \nu_{i,t} \notag \\
\text{QIX}_{i,q,t} &= \beta_{0} + \beta_{1}\text{R2000}_{i,t} + \beta_{2}\text{Rank}_{i,t} + \beta_{3}\text{R2000}_{i,t}\times\text{Rank}_{i,t} + \varepsilon_{i,t}
\end{align*}
\label{table:iortax}
\end{table}
\end{document}